\documentclass[12pt]{article}
\def\be {\begin{equation}}
\def\beq {\begin{equation}}
\def\ee {\end{equation}}
\def\feq {\end{equation}}
\def\ba {\begin{eqnarray}}
\def\ea {\end{eqnarray}}

\def\lb {\label}
\def\order{{\cal {O}}}
\def\bi {\begin{itemize}}
\def\ei {\end{itemize}}
\begin{document}
\def\bea{\begin{eqnarray}}
\def\eea{\end{eqnarray}}

\title{Near-horizon limit of the charged BTZ black hole and AdS$_{2}$
quantum gravity
}

\author{ Mariano Cadoni\footnote{email: mariano.cadoni@ca.infn.it }\,\,$^1$,
and Mohammad R. Setare \footnote{email: rezakord@ipm.ir}\,\,$^{2}$
\\
{$^1$\it \small Dipartimento di Fisica, Universit\`a di Cagliari and
INFN, Sezione di Cagliari }\\
{\small Cittadella Universitaria, 09042 Monserrato, Italy}\\
{$^2$ \it \small Department of Science,  Payame Noor University.
Bijar, Iran}\\}
\vfill

\maketitle

{\bf Abstract.}
We show that the 3D charged Banados-Teitelboim-Zanelli (BTZ) black hole solution interpolates
between  two different 2D AdS spacetimes: a near-extremal, near-horizon 
AdS$_{2}$ geometry with constant dilaton and $U(1)$ field and an
asymptotic AdS$_{2}$ geometry with a linear dilaton. Thus, the
charged BTZ
black hole can be considered as  interpolating between the two
different  formulations proposed until now for AdS$_{2}$ quantum
gravity. In both cases the theory  is the  chiral half of a
2D CFT and describes, respectively,  Brown-Hennaux-like
boundary deformations  and near-horizon excitations.
The  central charge $c_{as}¥$ of the  asymptotic CFT
is determined by 3D Newton constant $G$ and the AdS length $l$, $c_{as}¥=3l/G$,
whereas that of
the near-horizon CFT also depends on the $U(1)$ charge $Q$,
$c_{nh}\propto l Q/\sqrt G$.

\section{Introduction}
Quantum gravity in low-dimensional  anti-de Sitter(AdS)
spacetime has  features that make it peculiar with
respect to
the higher-dimensional cases. For $d=2,3$ the theory is a conformal
field theory (CFT) describing (Brown-Hennaux-like)
boundary   deformations and has a central charge determined
completely by Newton
constant and the AdS length \cite{Brown:1986nw,Cadoni:1998sg,
NavarroSalas:1999up,Maloney:2007ud}.
Conversely, in  $d>4$, quantum gravity in AdS spacetimes should admit
a near-horizon description in terms  of BPS solitons and D-brane excitations, whose
low-energy limit is an $U(N)$ gauge theory \cite{Strominger:1996sh,
Maldacena:1997re,Aharony:1999ti}.

The difference between these two descriptions is particularly evident
in their application for computing the entropy of non-perturbative
gravitational configurations such as black holes, black branes and
BPS states.  Brown-Hennaux-like boundary excitations have been used with
success to give a microscopically explanation to
entropy of the BTZ black hole and of two-dimensional (2D) AdS (AdS$_{2}$)
black holes \cite{Strominger:1997eq,Cadoni:1998sg}. On the other hand,
D-brane excitations account
correctly for the entropy of extremal and near-extremal
Reissner-Nordstrom black holes in higher dimensions \cite{Strominger:1996sh}.

Moreover,  the status of the  AdS$_{2}$/CFT$_{1}$ correspondence
\cite{Cadoni:1998sg,NavarroSalas:1999up,Strominger:1998yg,
Cadoni:2000gm,Hartman:2008dq,Sen:2008yk,
Gupta:2008ki} remains  still enigmatic. The dual
CFT$_{1}$ has been identified both as a conformal  mechanics and as
a chiral half of a 2D CFT.
Progress towards a better understanding of the relationship
between low- and higher-dimensional AdS/CFT correspondence has been
achieved in Ref. \cite{Hartman:2008dq}. It has been shown that quantum gravity on
AdS$_{2}$
with  constant  electromagnetic (EM) field and dilaton can be described by the chiral half of a
twisted CFT with central charge proportional to the square of the EM 
field.

On the other hand, there is another formulation  of  AdS$_{2}$ 
quantum gravity,
which uses  Brown-Hennaux-like boundary states in a 2D AdS
spacetime endowed with a linear dilaton \cite{Cadoni:1998sg}.
Also in this case
the Hilbert space
of the theory falls into the representation of a chiral half of a
CFT, but the central charge is proportional to the inverse of 2D newton constant.
The results of Ref. \cite{Hartman:2008dq} raise the question about the relationship
between the two different realizations of AdS$_{2}$ quantum gravity.

In this paper we show that a bridge between these two formulations
is  three-dimensional (3D) AdS-Maxwell gravity. We find that the charged BTZ black hole
admits two  limiting  regimes (near-horizon and asymptotic) in which
the black hole is described by a 2D  Maxwell-dilaton theory of
gravity. In the near-horizon, near-extremal regime the black hole is
described by AdS$_{2}$ with a constant dilaton and $U(1)$ field.
In the asymptotic regime the BTZ black hole is described
by  $AdS_{2}$ with a linear dilaton background and $U(1)$ field
strength $F_{tr}=Q/r$.

Both regimes are in
correspondence with a CFT$_{1}$, which can be thought as the chiral
half of a 2D CFT. The central charge of the near-horizon CFT is
proportional to the electric charge $Q$ of the BTZ black hole
$c_{nh}=(3k/4)\sqrt{\pi/G}lQ$ where $k$ is the level of the $U(1)$
current. The central charge of the asymptotic
 CFT is  determined completely by 3D Newton constant $G$
and the AdS length $l$: $c_{as}=3l/G$.

We can therefore think of the charged BTZ black hole as an
interpolating solution between the near-extremal, near-horizon behavior typical of
BPS-like solutions in higher dimensions (e.g. Reissner-Nordstrom black hole
solutions in four and five dimensions) and the asymptotic behavior
typical of Brown-Henneaux-like states.

This paper is organized as follows. In section 2 we review briefly
 the features of the charged BTZ black hole. In sect. 3 we
 investigate the two limiting regimes, namely the near-horizon limit
 and the asymptotic $r\to \infty$ limit. In sect. 4 we describe the
 dimensional reduction from three to  two spacetime dimensions. In sect. 5 we
 investigate the CFTs that describe the two different regimes and
 calculate the corresponding central charges.
 Finally, in section 6 we present our conclusions.

\section{The charged BTZ black hole }
The  charged BTZ black hole solutions are a generalization of the
well-known  black hole solutions in $(2+1)$ spacetime dimensions derived by Banados,
Teitelboim  and Zanelli
\cite{Banados:1992wn,Banados:1992gq}.

They are derived from a three-dimensional theory of gravity \be
I=\frac{1}{16\pi G}\int d^{3}x \sqrt{-g_{(3)}¥}\,
(R+\frac{2}{l^{2}}-4\pi G F_{\mu\nu}F^{\mu\nu}) \label{ac2}, \ee
where $G$ is 3D Newton constant, $\frac{1}{l^2}$ is the
cosmological constant ($l$ is the AdS-length) and $F_{\mu\nu}$ is
the electromagnetic  field strength. We  consider the BTZ black
hole with zero angular momentum  and  use  the conventions of  Ref.
\cite{Cadoni:2007ck}.

Electrically charged black hole solutions of the action (\ref{ac2})
are characterized by
the $U(1)$ Maxwell field \cite{Banados:1992wn,Martinez:1999qi},
\be\label{maxw}
F_{tr}=\frac{Q}{r},\label{metric3}
 \ee
where $Q$ is the electric charge.
The 3D  line element is given by
\be ds^2_{3}¥ =- f(r)dt^2
+ f^{-1}dr^{2}¥+r^2d\theta^2,
\label{metric}\ee
with metric function:
\be \label{charged}
f(r)=-8GM+\frac{r^2}{l^2}
-8\pi GQ^2 \ln (\frac{r}{l}), \hspace{0.5cm}
\ee
where $M$ is the Arnowitt-Deser-Misner (ADM) mass,
and $-\infty<t<+\infty$, $0\leq r<+\infty$,
 $0\leq \theta \leq2\pi$.
 The  black hole has one inner ($r_{-}$) and outer
($r_{+}$) , one or no horizons   depending
on whether
\be \label{ev}\Delta=8GM-4\pi G Q^2[1-2\ln(2Q\sqrt{\pi
G})] \ee
is greater than, equal to or less than zero,
respectively.
 Although these solutions for
$r\to\infty$ are asymptotically AdS,
they have a power-law curvature singularity at $r=0$,
 where $R\sim (8\pi GQ^2)/r^2$. This $r\to 0$ behavior of the
 charged BTZ black hole has to be compared with that of the uncharged
 one, for which $r=0$ represents just a singularity of the causal
 structure.

The Hawking temperature $T_H$ associated with the outer  black hole
horizon is
\be \label{tem}T_H=\frac{r_+}{2\pi l^2}-\frac{2 G
Q^2}{r_+}.\ee
According to the Bekenstein-Hawking formula, the
thermodynamic entropy of a black hole is proportional to the area $A$ of
the outer event horizon,  $S=\frac{A}{4G}$. For the charged BTZ black hole we have
\be \label{en}
S=\frac{\pi r_+}{2G}= \frac{\pi l}{G} \sqrt{2GM+ 2\pi
GQ^{2}\ln\frac{r_{+}}{l}}.\ee

\section{The near-horizon limit}
We are interested in the near-horizon, near-extremal behavior
of the solution (\ref{metric}). It is  well known  that in this
 regime asymptotically flat charged black holes in $d\ge
4$ dimensions are described by a $AdS_{2}\times S^{d-2}$ geometry,
i.e a Bertotti-Robinson spacetime. The flux of the EM field stabilize
the radius of the transverse sphere, so that in the near-horizon,
near-extremal limit it  becomes constant and given in terms of the EM
charge. Let us show that this is also the case for the charged BTZ
black hole.

The extremal limit $r_{+}=r_{-}=\gamma$ of the BTZ black hole
 is characterized by
$\Delta=0$ in Eq. (\ref{ev}), so that  $\gamma$ is a double zero of
the metric function (\ref{charged}):
\be\lb{f16}
\gamma= 2\sqrt{\pi G} Q l.
\feq

In order to describe the near-horizon, near-extremal limit of our
three-dimensional solution  we perform a translation of the radial
coordinate $r$,
\be\lb{f17}
r=\gamma+x,
\feq
and expand both the metric function (\ref{metric}) and $U(1)$ field
(\ref{maxw}) in powers of $x$. We get after some manipulations
\beq\lb{f17a}
f(x)= \frac{2}{l^{2}} x^{2}- 8G\Delta M +O(x^{3}),\quad F_{tx}=
\frac{1}{2\sqrt{\pi G}\,l}+ O(x),
\feq
where $\Delta M= M-M(\gamma)= M- \pi Q^{2}( \frac{1}{2}-
\ln(2Q\sqrt{\pi G}))$ is the  mass above extremality.
In the near-horizon, near-extremal limit the topology of the 3D
solution factorize as   $AdS_{2}\times S^{1}$ and the geometry
becomes that of 3D Bertotti-Robinson spacetime,
\be ds^2_{(3)}¥ =- ( \frac{2}{l^{2}} x^{2}- 8G\Delta M)dt^2
+ ( \frac{2}{l^{2}} x^{2}- 8G\Delta M)^{-1}¥dx^{2}¥+\gamma^2d\theta^2,
\quad F_{tx}=\frac{1}{2\sqrt{\pi G}l}.
\label{metricnh}\ee
The mass of the excitations above extremality can be also expressed in
terms of $\Delta r_{+}=r_{+}- \gamma$. Up to order three in $\Delta
r_{+}$ we have
\be\lb{exe}
\Delta M= \frac{\Delta r_{+}^{2}¥}{4G l^{2}}.
\feq
The near-horizon, extremal limit of the 3D charged AdS black hole is therefore very
similar to that of its higher-dimensional, asymptotically flat, cousins  such as the
Reissner-Nordstrom solution in four and five dimensions.
In particular, our 3D solution shares with them the thermodynamical
behavior. From Eqs. (\ref{tem}). (\ref{en}), (\ref{f17}) one easily
finds that  the extremal charged BTZ black hole is a state of zero
temperature and constant entropy
\be\lb{enex}
S_{(ext)}=\frac{\pi \gamma}{2G}= \pi \sqrt{\frac{\pi}{G}} Ql.
\feq
For small excitations near extremality we get using (\ref{exe})
\beq\lb{entropynh}
S_{ne}= \frac{\pi \gamma}{2G}+\pi  \frac{\Delta r_{+}}{2G}= \frac{\pi
\gamma}{2G}+
\pi l\sqrt{\frac{\Delta M}{G}}.
\feq

\subsection{The asymptotic  $r\to \infty$ limit}
It is also interesting to discuss briefly the asymptotic $r\to \infty$
limiting case of
the 3D solution (\ref{metric}) and its relationship with the
near-horizon solution (\ref{metricnh}). In the $r\to\infty$ limit
the metric describes
3D AdS spacetime, whereas the $U(1)$ field  goes to zero as
$1/r$. As we shall see in detail in the next section also in this regime
the 3D solution admits an effective description in terms of $AdS_{2}$
endowed with a linear varying dilaton. The dilaton parametrizes the
radius of the transverse one-sphere, which in the $r\to\infty$ limit
diverges.
We can therefore think of the full charged BTZ solution
(\ref{metric}) as a 3D spacetime interpolating between two regimes admitting an
effective description in terms of AdS$_{2}$.

\section{Dimensional reduction of the charged BTZ black hole}
The two limiting regimes of the BTZ black hole can be described by an
effective 2D Maxwell-Dilaton gravity model.
In order to find this 2D description, we parametrize the radius of the
$S^{1}$ sphere in the 3D solution (\ref{metric}) with a scalar field (the dilaton)
$\phi$:
\beq\lb{dr}
ds_{(3)}^{2}= ds_{(2)}^{2}+l^{2}\phi^{2}d\theta^{2}.
\feq
where  $ds_{(2)}$ is the line element of the 2D  sections of the 3D
spacetime covered by the $(t,r)$ coordinates and $\phi$ is a function
of $t,r$ only.
We will  consider only electric configurations for the
3D maxwell field, i.e we use for $F_{\mu\nu }$ the ansatz
\beq\lb{anf}
F_{t\theta}=F_{r\theta}=0.
\feq

Using Eqs. (\ref{dr}) and (\ref{anf}) into the 3D action (\ref{ac2})
one obtains, after defining  the rescaled dilaton
$\eta= (l/4G) \phi$, the dimensionally reduced
 2D action,
 \beq\lb{2da}
 I=\frac{1}{2}\int d^{2}x \sqrt{-g¥}\,
\eta \left(R+\frac{2}{l^{2}}-4\pi G F^{2}\right) . \ee
The field equation stemming from the action (\ref{2da}) are
\bea\lb{fe}
&&R+\frac{2}{l^{2}}-4\pi GF^{2}=0\nonumber\\
&&\nabla_{\mu}¥(\eta F^{\mu\nu}¥)=0\nonumber\\
&&-\nabla_{\mu}\nabla_{\nu}\eta + \left[ \nabla^{2}\eta
-\frac{\eta}{l^{2}} + 2\pi G \eta F^{2}\right]g_{\mu\nu }= 8\pi G
\eta F_{\mu\beta} F_{\nu}^{\beta}. \eea It is important to notice that
the field equations are invariant under rescaling of the dilaton by
a constant. This constant mode of the dilaton is therefore
classically undetermined but it can be fixed by matching the 2D with
the 3D solution.

The field equations (\ref{fe}) admit two classes of solutions  whose metric
part is
always a 2D AdS spacetime: $1)$ AdS$_{2}$ with linear dilaton and
with electric field which vanishes asymptotically (corresponding to
the asymptotic $r\to \infty$ regime of the charged BTZ black hole);  $2)$
AdS$_{2}$ with constant dilaton and electric field
(corresponding to the near-horizon limit of the BTZ black hole).
Let us discuss separately  these solutions.

\subsection{ AdS$_{2}$ with a linear dilaton}

This solution of the field Eqs. (\ref{fe}) is just the 3D solution
(\ref{metric}) written in a two-dimensional form, \beq\lb{ld}
ds^{2}= -f(r) dt^{2}+ f^{-1}(r) dr^{2}, \quad
F_{\mu\nu}=\frac{Q}{r}\epsilon_{\mu\nu},\quad
\eta=\bar\eta_{0}\frac{r}{l} \feq 
where  $f(r)$ has exactly the same
form as given by Eq. (\ref{charged}), $Q$ is the electric charge
and $\bar\eta_{0}$ is an integration constant  related to the scale
symmetry of the  2D field equations. The integration constants
appearing in Eq. (\ref{ld}) (thus defining the physical parameters
of the 2D black hole)  can be easily identified in terms of the
physical parameters of  the BTZ black hole. The charge $Q$  and mass
$M$ of the 2D black hole are the same as those of the BTZ black
hole. The constant  $\bar \eta_{0}$  is determined
by the ansatz (\ref{dr}), \beq\lb{dil} \bar \eta_{0}=\frac{l}{4G}.
\feq With this identification also the temperature and entropy of
the 2D black hole match exactly  those for the 3D black hole given
by Eqs. (\ref{tem}) and (\ref{en}). For instance, the  entropy of
the 2D black hole is determined by the value of the dilaton on the
horizon, \beq S= 2\pi\eta_{horizon}, \feq which after using Eqs.
(\ref{ld}) and (\ref{dil}) reproduces exactly Eq. (\ref{en}).

\subsection{ AdS$_{2}$ with constant dilaton and electric field}
One can easily realize that the field equations (\ref{fe})  admit a solution
describing AdS$_{2}$
with   constant dilaton and electric field.
The constant
value of the dilaton, which is   not fixed by the 2D field equations,
is determined by the ansatz (\ref{dr}),
\beq\lb{eta}
\eta_{0}= \frac{l}{2} \sqrt{\frac{\pi}{G}}\, Q.
\feq
In order to have the usual normalization of the electric field and to
make contact with the model investigated in Ref. \cite{Hartman:2008dq},
it is necessary  to perform a Weyl transformation of the metric and
a rescaling of the $U(1)$ field strength:
\beq\lb{wt}
g_{\mu\nu}= \frac{\eta}{\eta_{0}} \bar g_{\mu\nu},\quad F_{\mu\nu}=
\frac{l}{2 \sqrt{2\pi G\eta_{0}}} \bar F_{\mu\nu}.
\feq
After this transformation the 2D action (\ref{2da}),
modulo total derivatives, becomes
\beq\lb{2da1}
 I=\frac{1}{2}\int d^{2}x \sqrt{-\bar g¥}\left[\,
\eta \left(R(\bar g)+\frac{(\partial\eta)^{2}}{\eta}+\frac{2\eta}{l^{2}\eta_{0}¥}\right)-
\frac{l^{2}}{2} \bar F^{2}\right]. \ee
The field equations stemming from this action allow for a solution
describing  $AdS_{2}$
with constant dilaton and electric field, which is the
dimensional
reduction of the near-horizon solution (\ref{metricnh})
\bea\lb{nh}
ds^{2}&=& -( \frac{2}{l^{2}}x^{2}-k^{2}) dt^{2}+ (
\frac{2}{l^{2}}x^{2}-k^{2})^{-1}¥ dx^{2}, \quad
\bar F_{\mu\nu}=2E \epsilon_{\mu\nu},\nonumber\\
\quad \eta&=&2l^{4}E^{2},\quad  E^{2}= \frac{1}{4 l^{3}} 
\sqrt{\frac{\pi}{G}}\, Q,
\eea
where we have used Eq. (\ref{eta}) and $k^{2}=8G\Delta M$.

Following Ref. \cite{Hartman:2008dq} we can linearize the term
quadratic in the $U(1)$ field strength by introducing in the action an
auxiliary field $h$,
\beq\lb{action}
I=\frac{1}{2}\int d^{2}x \sqrt{-\bar g¥}\,
\left[\eta \left(R(\bar g)+\frac
{(\partial\eta)^{2}}{\eta}+\frac{2\eta}{l^{2}\eta_{0}¥}\right)-
\frac{h^{2}}{l^{2}}+ h\epsilon^{\mu\nu}\bar F_{\mu\nu}\right]. \feq
The field equations for $h$ give
\beq\lb{hs}
h=\frac{l^{2}}{2}\epsilon^{\mu\nu}\bar F_{\mu\nu}=-
2E l^{2}¥.
\feq

\section{ Conformal symmetry and central charges}

In view of the AdS/CFT correspondence, the existence of two limiting
AdS$_{2}$ configurations for the charged BTZ
black hole  imply the  duality of the gravitational configuration
with two different CFTs.
Both  CFTs have been  already investigated in the literature and in
both of them the conformal transformations appear as a subgroup of the
2D diffeomorphisms. However, they differ in the way the central charge
of the CFT is generated.
The CFT associated with  the $r\to \infty$ limit, corresponding to AdS$_{2}$
with a linear
dilaton has been investigated in Ref. \cite{Cadoni:1998sg}. In this case the
central charge of the CFT is generated by the breaking of the $SL(2,R)$
isometry  of the AdS$_{2}$ background due to the non-constant dilaton
\cite{Cadoni:2000ah}.

The CFT associated with the near-horizon limit, corresponding to
AdS$_{2}$ with a constant electric and dilaton field  has been
investigated in Ref. \cite{Hartman:2008dq}. In this case the central
charge of the CFT is  generated by the boundary conditions for the EM
vector potential. We will discuss the two cases separately.

\subsection{The $r\to\infty$ asymptotic CFT}
In this case the conformal algebra is generated by the group of
asymptotic symmetries (ASG)  of AdS$_{2}$ along the lines of Ref.
\cite{Cadoni:1998sg,Cadoni:2007ck}.  The calculations of Refs. \cite{
Cadoni:1998sg} can be easily extended to the theory
described by the action (\ref{2da}). The only difference is the
presence of the $U(1)$ field, which however, as explained in Ref
\cite{Cadoni:2007ck} for the case of 3D gravity, does not change
neither  the conformal algebra, which is always given by a chiral
half of the Virasoro algebra, nor the value of the central charge.

The $r\to\infty$ boundary conditions for the  fields,  which are
invariant under 2D diffeomorphisms generated by killing vectors
$\chi^{t}= l\epsilon(t) +\order(1/r^{2}),\chi^{r}= -l r\dot 
\epsilon(t)
+\order(1/r)$ are
\ba\lb{bc}
g_{tt}&=& -\frac{r^{2}}{l^{2}}+\order(\ln r),\quad
g_{tr}=\order(\frac{1}{r^{3}}),\nonumber\\
g_{rr}&=& \frac
{l^{2}}{r^{2}}+\order(\frac{\ln r}{r^{4}}),\quad \eta=\order(r),\quad
F_{tr}=\order(\frac{1}{r}).
\ea
Notice that we allow for deformations of the dilaton and EM field that
are of the same order of the background solution (\ref{ld}).
Although the boundary conditions (\ref{bc}) are invariant under the
action of the asymptotic symmetry group, the classical solution is not.
The linear dilaton and the $Q/r$ EM field break the isometry group of
AdS$_{2}$.  The breaking of the isometry group due to the linear
dilaton background produces a nonvanishing central charge in the
conformal algebra \cite{Cadoni:2000ah}. Conversely, the EM field  does not
contribute to the boundary charges, but only enters in the
renormalization of the  $L_{0}$ Virasoro operator \cite{Cadoni:2007ck}.

The generators of the conformal diffeomorphisms close in the Virasoro algebra
\beq\label{va}
[L_{m},L_{n}]=(m-n) L_{m+n}+ \frac{c}{12}(m^{3}-m)\delta_{m+n\,0}.
\feq
The central charge $c$ can be computed using a canonical realization of
the ASG  along the lines of Refs. \cite{Cadoni:1998sg,Cadoni:2000gm,
Cadoni:2007ck}. One has
\beq\lb{cc}
c= 12\bar \eta_{0}=\frac{3l}{G}.
\feq
where we have used  Eq. (\ref{dil}).

The eigenvalue $l_{0}$ of the Virasoro operator $L_{0}$ is related to
the black hole mass $M$. Analogously to the 3D case, the asymptotic
expansion (\ref{bc}) gives divergent contributions to the boundary
charges. A renormalization procedure \cite{Martinez:1999qi,Cadoni:2007ck})
allows for the
definition of  renormalized boundary charges and in particular of  
renormalized mass $M_{0}(r_{+})$, which has to be interpreted as the
total energy (gravitational and electromagnetic) inside the horizon
$r_{+}$.
\beq\lb{rm}
M_{0}(r_{+})= M+\pi Q^{2}\ln(\frac{r_{+}}{l}).
\feq
The eigenvalue of  $L_{0}$ is  therefore
\beq\lb{lzero}
l_{0}=lM_{0}(r_{+})= l [M + \pi Q^{2} \ln(\frac{r_{+}}{l})]
\feq

\subsection{The near-horizon CFT}
The  2D action (\ref{action}) can be recast in the form of a twisted 2D CFT
in which  a central term in the Virasoro algebra is generated by
boundary conditions for the $U(1)$ vector potential $A_{\mu}$, along the
lines of Ref. \cite{Hartman:2008dq}.

Using a conformal and Lorentz gauge respectively, we fix the
diffeomorphisms and $U(1)$ gauge freedom, \beq\lb{gauge} ds^{2}=
-e^{2\rho}dx^{+}dx^{-},\quad \partial_{\mu}A^{\mu}=0, \feq the
action (\ref{action}) becomes up to total derivatives \beq\lb{h2}
I=\frac{1}{2}\int d^{2}x \left( -4 \partial
_{-}\eta\partial_{+}\rho+ \frac{\eta}{l^{2}\eta_{0}} +2
\frac{\partial _{-}\eta\partial_{+}\eta}{\eta}- \frac{h^{2}}{2l^{2}}
+ 4 \partial _{-}h\partial_{+}a\right), \feq where we have  used the
fact that in the gauge (\ref{gauge}) $A_{\mu}¥$  can be given in terms of a
scalar $a$,  $A^{\mu}=\epsilon^{\mu\nu}\partial_{\nu}¥a$.

As usual for gauge-fixing the classical field equations
stemming from the action (\ref{h2}) must be supported by constraints,
\bea\lb{h4}
&&T_{\pm\pm}=\frac{2}{\sqrt{- g}}\frac {\delta I}{\delta g^{\pm\pm}}=
-2\partial_{\pm}\eta\partial_{\pm}\rho + \partial_{\pm}\partial_{\pm}\eta
- \eta^{-1}¥\partial_{\pm}\eta\partial_{\pm}\eta+
2\partial_{\pm}h\partial_{\pm}a=0,\\
&& J_{\pm}=2\frac {\delta I}{\delta A^{\pm}}=\pm 2\partial_{\pm} h=0.
\eea The stress-energy tensor $T_{\pm\pm}$ and the $U(1)$ current
$J_{\pm}$ are (classically) holomorphic conserved and generate,
respectively, residual conformal diffeomorphisms and gauge
transformations.

In the conformal gauge the vacuum AdS$_{2}$ solution (\ref{nh})
becomes
\beq\lb{h5}
ds^{2}= -2l^{2}\frac{dx^{+}dx^{-}}{(x^{+}-x^{-})^{2}¥},\quad
A_{\pm}= \frac{El^{2}}{2\sigma},
\feq
where $\sigma= (1/2)(x^{+}-x^{-})$ and $h,\eta,E$ are given by Eqs.
(\ref{nh}), (\ref{hs}).

Because the dilaton is constant one naively expects that we are
dealing with pure  2D quantum gravity, which is known to be
described by a CFT with vanishing central charge
\cite{Distler:1988jt}. However, it has been shown in
\cite{Hartman:2008dq} that the boundary conditions for the $U(1)$
vector potential at the $\sigma=0$ conformal boundary of AdS$_{2}$,
$A_{\sigma}|_{\sigma=0}=0$, is not preserved by conformal
diffeomorphisms generated by $\chi^{+}(x^{+})$ and $\chi^{-}(x^{-})$.
It must be accompanied by a gauge transformation $\omega^{+}(x^{+})+
\omega^{-}(x^{-})$,  which in the case under consideration is given
by \beq\lb{h7}
\omega^{\pm}=\mp\frac{l^{2}E}{2}\partial_{\pm}\chi^{\pm}. \feq
Moreover,  the requirement the  boundary remains at $\sigma=0$
determines  a chiral half of the conformal diffeomorphisms in terms
of the second half. The  resulting conformal symmetry  can be
realized using
 Dirac brackets. Conformal transformations are generated
by the improved stress-energy tensor \beq\lb{h15} \tilde
T_{--}=T_{--}- \frac{El^{2}}{2}\partial_{-}J_{-}. \feq
Expanding in Laurent  modes and using
the transformation law of the improved stress-energy tensor
\beq\lb{h16}
\delta_{\chi}\tilde T_{--}= \chi^{-}\partial_{-}\tilde T_{--}+2\partial_{-}
\chi^{-} \tilde T_{--}+ \frac{c}{12}\partial_{-}^{3}\chi^{-},
\feq
where we allow for the existence of an anomalous term, one finds  that
the operators $\tilde L$
span the Virasoro algebra (\ref{va}).
The transformation law of the original $T_{--}$ is anomaly-free, but
that of the current $J_{-}$ may have an anomalous term proportional to
its level $k$ \cite{Hartman:2008dq},
\beq\lb{h18}
\delta_{\omega}J_{-}= k\partial_{-}\omega^{-}.
\feq
This allows us  to compute the central charge $c$ of the Virasoro
algebra,
\beq\lb{h20}
c=3k E^{2}l^{4}=\frac{3}{4} k\sqrt{\frac{\pi}{G}}\, l Q.
\feq

\section{ Conclusion}
Using the results of the previous section we can reproduce the
entropy of the 2D AdS black hole (and  the entropy of the chraged BTZ black
hole)  by calculating the density of states $\rho(l_{0})$ of the CFT with
a given eigenvalue $l_{0}$. In the semiclassical limit $c>>1$
and for large $l_{0}$ we have  Cardy formula,
\beq\lb{cardy}
S=\ln \rho(l_{0})= 2\pi \sqrt{\frac{cl_{0}}{6}}
\feq
Using Eqs (\ref{cc}) and (\ref{lzero}) we reproduce exactly
the black hole entropy (\ref{en}).

In principle, one should also be able to reproduce the entropy of the
near-extremal black hole (\ref{entropynh}) using a similar
procedure for the
near-horizon twisted CFT.  However, naive application of Cardy
formula in this case is not possible.
The 2D solution (\ref{nh}) has
zero mass. Although the spacetime has an horizon and
we may assign to it an Hawking temperature the 2D solution
cannot be interpreted as a black hole.
Being characterized by a constant dilaton and
EM field, there is nothing to prevent maximal extension
of the spacetime beyond the horizon  to recover full AdS$_{2}$.
Thus, the horizon is not an event horizon but  has to be
seen as an acceleration horizon.
This is not the case of AdS$_{2}$ endowed with a linear dilaton.
The dilaton is a  non-constant scalar  and its inverse gives 2D Newton constant.
The point  $r=0$  has to be considered
a spacetime singularity and the horizon in Eq. (\ref{ld}) an event
horizon \cite{Cadoni:1994uf}.

The vanishing of the mass for the near-horizon solution implies
$l_{0}=0$, which in turn implies a vanishing entropy for the untwisted
near-horizon CFT. However, to calculate the density of states for
the twisted CFT we have to use in the cardy formula (\ref{cardy})
the eigenvalues of $\tilde L_{0}$, $\tilde l_{0}$, instead of that of 
$L_{0}$. $\tilde l_{0}$
may still be non zero. Calculation of $\tilde l_{0}$ requires
careful analysis of the CFT spectrum  and detailed knowledge of the
effect of the twisting on the Hilbert space of the 2D CFT.

{\bf Acknowledgements}\\
We thank G. D'Appollonio  for discussions and valuable comments.

\end{document}